\documentclass[%
 aip,
 amsmath,amssymb,
reprint,%
]{revtex4-1}

\usepackage{graphicx}% Include figure files
\usepackage{dcolumn}% Align table columns on decimal point
\usepackage{bm}% bold math
\usepackage[utf8]{inputenc}
\usepackage[T1]{fontenc}
\usepackage{mathptmx}
\usepackage{etoolbox}
\usepackage{listings}
\usepackage{xcolor}
\usepackage{multirow}
\usepackage[version=4]{mhchem}
\usepackage{booktabs}
\usepackage{cancel}

\definecolor{codegreen}{rgb}{0,0.6,0}
\definecolor{codegray}{rgb}{0.5,0.5,0.5}
\definecolor{codepurple}{rgb}{0.58,0,0.82}
\definecolor{backcolour}{rgb}{0.95,0.95,0.92}

\lstdefinestyle{mystyle}{
  backgroundcolor=\color{backcolour}, commentstyle=\color{codegreen},
  keywordstyle=\color{magenta},
  numberstyle=\tiny\color{codegray},
  stringstyle=\color{codepurple},
  basicstyle=\ttfamily\footnotesize,
  breakatwhitespace=false,         
  breaklines=false,                 
  captionpos=b,                    
  keepspaces=true,                 
  numbers=left,                    
  numbersep=1pt,                  
  showspaces=false,                
  showstringspaces=false,
  showtabs=false,                  
  tabsize=1,
  columns=fixed,
  basicstyle=\fontsize{7pt}{8 pt}\selectfont,
}

\newcommand{\piplayer}{\texttt{PIPLayer} }

\DeclareUnicodeCharacter{0394}{\ensuremath{\Delta}}

\DeclareMathOperator{\pipvec}{\boldsymbol{\Phi_{\text{PIP}}}}
\DeclareMathOperator{\pipvecaniso}{\boldsymbol{\Phi_{\text{PIP}}^{\text{aniso}}}}
\DeclareMathOperator{\vx}{\mathbf{x}}
\DeclareMathOperator{\vz}{\mathbf{z}}
\DeclareMathOperator{\vecr}{\mathbf{r}}
\DeclareMathOperator{\vgamma}{\bar{\boldsymbol{\gamma}}}
\DeclareMathOperator{\npip}{n_{p}}
\DeclareMathOperator{\na}{n_{a}}

\DeclareMathOperator{\fnn}{f_{\text{NN}}}
\DeclareMathOperator{\fpipnn}{f_{\text{PIP-NN}}}
\DeclareMathOperator{\fmono}{f_{\text{mono}}}
\DeclareMathOperator{\fpoly}{f_{\text{poly}}}
\DeclareMathOperator{\lambdazero}{\lambda_0}
\DeclareMathOperator{\vlambda}{\boldsymbol{\lambda}}

\makeatletter
\def\@email#1#2{%
 \endgroup
 \patchcmd{\titleblock@produce}
  {\frontmatter@RRAPformat}
  {\frontmatter@RRAPformat{\produce@RRAP{*#1\href{mailto:#2}{#2}}}\frontmatter@RRAPformat}
  {}{}
}%
\makeatother
\begin{document}

% \preprint{AIP/123-QED}

\title{MOLPIPx: an end-to-end differentiable package for permutationally invariant polynomials in Python and Rust}
% Force line breaks with \\
\author{Manuel S. Drehwald}
\thanks{These authors contributed equally.}
\affiliation{Department of Computer Science, University of Toronto, Toronto, ON, Canada}%Lines break automatically or can be forced with \\
\affiliation{Department of Chemistry and Chemical Biology, McMaster University, Hamilton, ON, Canada}
\author{Asma Jamali}%
\thanks{These authors contributed equally.}
\affiliation{School of Computational Science and Engineering, McMaster University, Hamilton, ON, Canada}
\affiliation{Department of Chemistry and Chemical Biology, McMaster University, Hamilton, ON, Canada}
\author{Rodrigo A. Vargas-Hernández}
\email{vargashr@mcmaster.ca}
\affiliation{Department of Chemistry and Chemical Biology, McMaster University, Hamilton, ON, Canada}
\affiliation{School of Computational Science and Engineering, McMaster University, Hamilton, ON, Canada}
\affiliation{Brockhouse Institute for Materials Research, McMaster University, Hamilton, ON, Canada}

% \date{\today}% It is always \today, today,
%              %  but any date may be explicitly specified
             
\begin{abstract}
In this work, we present MOLPIPx, a versatile library designed to seamlessly integrate Permutationally Invariant Polynomials (PIPs) with modern machine learning frameworks, enabling the efficient development of linear models, neural networks, and Gaussian process models. These methodologies are widely employed for parameterizing potential energy surfaces across diverse molecular systems. MOLPIPx leverages two powerful automatic differentiation engines—JAX and EnzymeAD-Rust—to facilitate the efficient computation of energy gradients and higher-order derivatives, which are essential for tasks such as force field development and dynamic simulations. MOLPIPx is available at  \url{https://github.com/ChemAI-Lab/molpipx}.
\end{abstract}

\maketitle

% ----------------------------------------------------------------------------------------
\section{\label{sec:intro}Introduction}
The construction of potential energy surfaces (PESs) has been a long challenge in the computational chemistry community\cite{pes:bowman:2004,pes:bowman:2014,pes:bowman1999,pes:bowman:2007,pes:bowman:2012,pes:bowman2003,pes:bowman:2003:2,pes:bowman:2015,pes:bowman:2017,pes:bowman:2006,pes:bowman:2002,pes:bowman:2014,pes:bowman:2003:3,pes:bowman:2024,pes:bowman:2014:2}. 
One of the most successful methodologies, permutationally invariant polynomials (PIPs), has a history of over 20 years and has inspired numerous reviews due to its success \cite{pes:bowman:2010, his_pip:dral:2020, his_pip:frohlking:2020, his_pip:jiang:2020, his_pip:jinnouchi;2020, his_pip:koner:2020, his_pip:poltavsky:2021, his_pip:tong:2020, his_pip:westermayr:2021}.
PIPs offer a unique approach to representing the structural information of molecules by ensuring invariance between permutations of like atoms within a molecule \cite{pip:bowman:2009, pip:bowman:2018}. This property is pivotal, as it allows models to effectively discern molecular features regardless of atom ordering, creating robust and efficient regression models \cite{pipgrad:bowman:2024,pip:bowman:2018}.
Given the flexibility of PIP models, they have also been adapted for the computation of the forces \cite{pipgrad:bowman:2002,pipgrad:bowman:2019,pipgrad:bowman:2020,piprevgrad:bowman:2022, pipgrad:bowman:2024} which has allowed their use in molecular dynamic simulations\cite{force:field:review2021,pipgrad:bowman:2019, md:murakami:2024,md:liu:2023,md:unke:2020,md:zhang:2023,Behler:2011}. 
% By using PIP-based models, researchers can accurately simulate the behavior of a molecular system, allowing for high-precision investigation of molecular interactions or reaction pathways\cite{Behler:2011}.

To date, more than 100 PIP-based PESs have been developed\cite{pip:bowman:2018}. 
Early examples of PIPs\cite{pesforce:bowman:2022} for PESs are \ce{CH5+}\cite{pes:bowman:2003}, \ce{H5+}\cite{pip:xie:2005}, and the chemical reaction \ce{H +CH4}\cite{pip:zhang:2006}. 
Subsequently, PIP models were developed for molecules with 7 or more atoms, including ethanol\cite{pip:bowman:2021}, nitromethane\cite{pip:wang:2015}, formic acid dimer\cite{pes:bowman:2016}, glycine\cite{pes:conte:2020}, N-methyl acetamide\cite{pip:bowman:2019:2,pip:bowman:2019,pipgrad:bowman:2020}, acetylacetone\cite{pes:bowman:2021}, and tropolone\cite{pip:bowman:2020}, as well as for the 4 water molecules\cite{pip:bowman2023}. 
In addition, PIPs have been used in the MB-pol-\emph{like} potentials~\cite{pip:babin:2014, pip:moberg:2016,pip:palos:2023,pip:palos:2023:2,pip:bull:2021,pip:riera:2023}, and other many-body type potentials for solvated molecules like \ce{CO2}/\ce{H2O}\cite{pip:riera:2020}. 
Recently, Bowman et al.\cite{pip:bowman:2024} published the first in a series of papers focused on alkanes of the form C$_n$H$_{2n+2}$, with a particular emphasis on \ce{C14H30}. 
A \texttt{ROBOSURFER} program has been used to study reactions like \ce{CH3Br + F-} using PIPs via automating the selection of new geometries, performing \emph{ab initio} computations, and iteratively refining the PES\cite{pip:gyoori:2019}. 
A method involving dictionary learning with "greedy" selection reduces the basis size with minimal accuracy loss and can automatically generate PIPs for complex systems using parallelization\cite{pip:moberg:2021,pip:moberg:2021:2}. 
Additionally, a permutationally invariant method using an "n-body" representation has been introduced for polyatomic molecules\cite{pip:koner:2020}. Several PESs, including \ce{N + N2}\cite{pip:varga:2021:2}, \ce{N + O2}\cite{pip:varga:2021} and \ce{N2 + N2}\cite{pip:li:2020}, have been studied using PIPs, with frameworks that remove "disconnected" terms, achieving purification and compaction\cite{pip:paukku:2013,pip:moberg:2021, pip:moberg:2021:2}. 
With the growth of computational power, PIPs have been recently adapted to study materials that rely on high-dimensional PESs \cite{pip:conte:2015, pip:Oord:2020, pip:allen:2021}. 

% big paragraph describing different systems where they have been used. 
Researchers have also explored different methodologies to effectively incorporate PIPs with modern machine learning (ML) models\cite{pesgp:Richard:2017, pip:balan:2022} like neural networks (NNs) and Gaussian Processes (GPs), for example, the \texttt{PES-Learn} software \cite{pes:Schaefer:2019}. 
With fairly successful results, researchers have shifted towards assessing the performance of these various models using specific molecules such as \ce{H3O2-}\cite{pipgrad:bowman:2024}, \ce{CH4}\cite{pipgrad:bowman:2019}, \ce{H3O+}\cite{Sotiris:2018, pesforce:bowman:2022} and \ce{C2H6O}\cite{pes:bowman:2022}, to mention a few\cite{pip:bowman:2019,pip:bowman:2019:2}.

PIPs were first integrated with neural networks to construct PESs for gas-phase molecules\cite{pipnn:review:Hua:2016} starting with triatomic\cite{ pipnn:Hua:JCP2013, pipnn:li:2014:4} (\ce{H + H2} and \ce{Cl + H2}) and tetratomic \cite{ pipnn:li:2013, pipnn:li:2013:6, pipnn:li:2013:7}(\ce{X + H2O -> HX + H2O} (X = H, F, O)) systems and later extended to several polyatomic systems\cite{pipnn:li:2014, pipnn:li:2014:2, pipnn:li:2014:3, pipnn:li:2014:4,pipnn:li:2015, pipnn:li:2015:2}. This framework has also been applied to molecular systems, like Criegee intermediates (\ce{CH2OO}\cite{pipnn:li:2014:5, pipnn:yu:2015} and dioxirane\cite{pipnn:li:2016}), acetylene-vinylidene\cite{pipnn:than:2014}, and the \ce{H2CC-} anion\cite{pipnn:guo:2015}. 
Furthermore, this approach was further extended to molecular surfaces like \ce{H2 + Cu(III)} and \ce{H2 + Ag(III)}\cite{pipnn:jiang:2014}.
For more details regarding the integration of PIPs in NN, we refer the reader to the review by S. Manzhos, Sergei et al.  \cite{pipnn:review:carrington:2021}.
Given the flexibility of NN, this combined framework has been extended to diabatic potential energy matrices\cite{pipnn:guan:2019, pipnn:xie:2018}. 
In addition, the combination of PIP, neural network, and $\Delta$-machine learning shows a robust and efficient approach for constructing highly accurate PESs across various molecular systems, from simple diatomic dissociations to complex multi-atom reactions\cite{deltaml:liu:2022}. 

The other successful integration of ML models with PIPs is when combined with Gaussian processes. This framework has been applied to various molecular systems, including \ce{H3O+}, \ce{OCHCO+}, \ce{(HCOOH)2}, and \ce{H2CO}/\ce{HCOH}, demonstrating its efficacy in terms of fitting accuracy, computational efficiency, and the ability to capture PES features such as stationary points and harmonic frequencies\cite{pipgp:vargas:2018}.

Given the extensive use of PIPs in materials science and computational chemistry, we present MOLPIPx, a flexible library that can seamlessly integrate PIPs with advancements in machine learning modelling, enhancing the simulation of a wide range of chemical systems. MOLPIPx takes advantage of two main automatic differentiation engines, JAX for the Python version and Enzyme-AD \cite{EnzymeAD:moses:2020, EnzymeAD:moses:2021, EnzymeAD:moses:2022} for the Rust version.

% ----------------------------------------------------------------------------------------
\section{\label{sec:models}Models}
\begin{figure}[h!]
    \centering
    \includegraphics[scale=0.3]{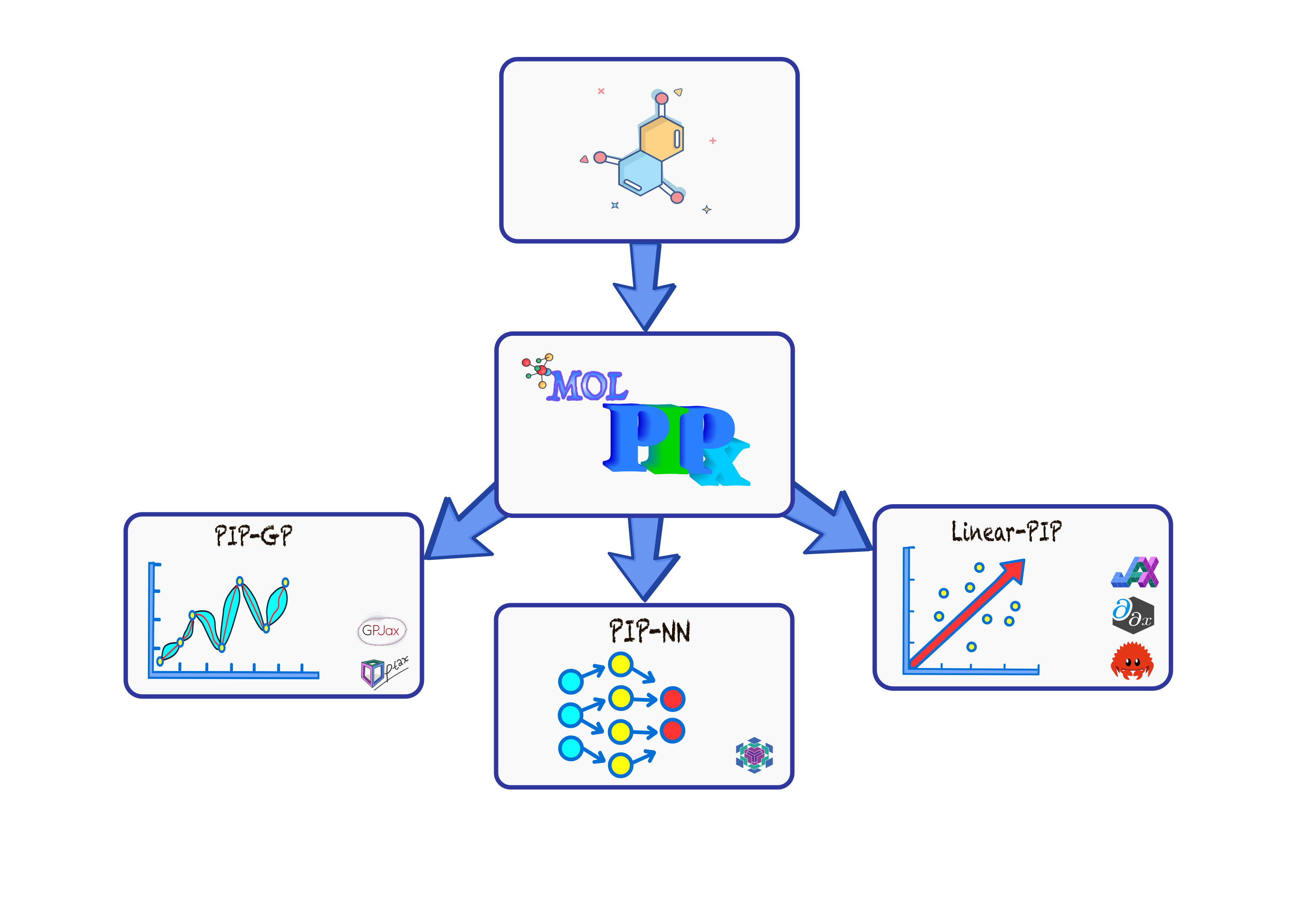}
    \vspace{-30pt}
    \caption{
Overview of the MOLPIPx package. For any model developed under MOLPIPx, the monomial and polynomial functions are constructed using the MSA algorithm and then translated into JAX or Rust, depending on the selected back-end. We provide some monomial and polynomial functions, see Table \ref{tab:molecules}. The input for all models is the Cartesian coordinates of a molecule, which are used to construct the PIP vector (Eq.~\ref{eqn:piplayer}). These vectors can then be integrated into other modern machine learning models to build PESs.
    \label{fig:pip_diagram}}
\end{figure}

This section describes the three main regression models in the MOLPIPx library. 
First, Sections \ref{sec:linear_pip} and \ref{sec:anisoPIP} introduce PIPs under the framework of function composition and its extension to linear models for PESs. This is followed by discussing PIPs' integration with ML algorithms, specifically neural networks and Gaussian Processes, Sections \ref{sec:pipnn} and \ref{sec:pipgp}. For more details regarding the PIP framework, we encourage the reader to consult Refs.~\cite{pip:bowman:2009, pip:bowman:2018, pip:bowman:2010, pip:bowman:2019, pip:bowman:2019:2}.
The general structure of MOLPIPx and its components is depicted in Fig.~\ref{fig:pip_diagram}.
\vspace{-20pt}
% The $\pipx$ library incorporates two main AD back-ends, jax and EnzymeAD. 
% For the jax back-end, the\texttt{jax.value\_and\_grad} function is used to compute both energy and force simultaneously for any PES model. 

\subsection{\label{sec:pip} Permutationally Invariant Polynomial Representation}
The PIP representation for PES models can be conceptually described as a complex composition of multiple functions that transform the Cartesian coordinates of a molecule ($\vx$) into a structured vector ($\pipvec$), which is invariant to permutations of similar nuclei. 
The primary motivation for articulating $\pipvec$ as a sequence of function compositions is to facilitate its integration with modern automatic differentiation frameworks (AD)~\cite{ad:survey}. 
AD has been fundamental in the numerical ecosystem of deep learning libraries and, more recently, in computational chemistry simulations. \cite{ad:blondel2022, ad:dral:2024,ad:survey,ad:tamayo:2018,ad:tan:2023,ad:vargas:2023,ad:vargas:2024,ad:zhang:2024,ad:arrazola:2021,ad:Casares:2024,ad:dawid:2022,ad:kasim:2021,ad:schmidt:2019,ad:vargas:2021,ad:zhang:2022}.
By structuring the PIP vector, each component function can be seamlessly differentiated and combined with other machine learning algorithms; Sections \ref{sec:pipnn} and \ref{sec:pipgp}. 

The construction of $\pipvec$ in MOLPIPx begins with calculating the inter-nuclei distances ($\vecr$), which are then transformed into Morse variables ($\vgamma$)~\cite{pip:bowman:2009}. The subsequent steps involve computing the symmetrized monomials ($\vz_{\text{mono}}$) and polynomials ($\vz_{\text{poly}}$) using the $\fmono$ and $\fpoly$ functions. 
These groups of transformations ensure that the resulting representation is invariant under the permutation of identical nuclei, thus capturing the correct molecular symmetries. 
Mathematically, the $\pipvec$ vector can be defined as follows,
\begin{eqnarray}
\pipvec(\vx)=(\fpoly \circ \fmono \circ \gamma \circ d)(\vx), \label{eqn:piplayer}
\end{eqnarray}
where each component in this composition of functions represents a non-linear transformation defined as,
\begin{eqnarray}
\vecr &=& d(\vx), \quad \text{(inter-nuclei distances)} \label{eqn:distances} \\
\vgamma &=& \gamma(\vecr) = e^{-\lambda \vecr}, \quad \text{(Morse variables)} \label{eqn:morse} \\
\vz_{\text{mono}} &=& \fmono(\vgamma), \quad \text{(symmetrized monomials)} \label{eqn:mono} \\
\vz_{\text{poly}} &=& \fpoly(\vz_{\text{mono}}) \quad \text{(symmetrized polynomials)}. \label{eqn:poly}
\end{eqnarray}
In Eq.~\ref{eqn:morse}, $\lambda$ is a length-scale hyper-parameter, and the $\fmono$ and $\fpoly$ functions are constructed using the monomial symmetrization algorithm (MSA)\cite{pip:bowman:2010}.
\vspace{-15pt}

\subsection{\label{sec:linear_pip} Linear Permutationally Invariant Polynomial  Models}
Linear regression frameworks are the most commonly used strategies for building PIP models for PESs, and are defined as,
\begin{eqnarray}
y(\vx) = \mathbf{w}^\top \pipvec(\vx) = \sum_{i=1}^{\npip} w_i \phi_i(\vx),\label{eqn:pip_linear}
\end{eqnarray}
where $y$ represents the potential energy, $\mathbf{w}$ denotes the linear weights, and $\npip$ is the number of elements in $\pipvec$ given a polynomial degree $p$. 
The optimization of $\mathbf{w}$ is typically achieved using least-squares methods, and as detailed in Ref. \cite{pipgrad:bowman:2019}, it can be adapted to include the forces by incorporating the Jacobian of $\pipvec$ for each geometry ($\nabla_{\vx_{i}} \pipvec$) in the least-squares equation,
{\small
\begin{eqnarray}
     \begin{bmatrix}
\phi_0(\vx_{1}) & \phi_1(\vx_{1}) & \cdots & \phi_{\npip}(\vx_{1})  \\
\vdots & \vdots &  & \vdots \\
\phi_0(\vx_{N}) & \phi_1(\vx_{N}) & \cdots & \phi_{\npip}(\vx_{N})  \vspace{0.1cm}\\
\frac{\partial \phi_0(\vx_{1})}{\partial x_{1}} & \frac{\partial \phi_1(\vx_{1})}{\partial x_{1}} & \cdots & \frac{\partial \phi_{\npip}(\vx_{1})}{\partial x_{1}}  \\
\vdots & \vdots &  & \vdots \\
\frac{\partial \phi_0(\vx_{1})}{\partial x_{3\na}} & \frac{\partial \phi_1(\vx_{1})}{\partial x_{3\na}} & \cdots & \frac{\partial \phi_{\npip}(\vx_{2})}{\partial x_{3\na}}  \\
\vdots & \vdots &  & \vdots \\
\frac{\partial \phi_0(\vx_{N})}{\partial x_{1}} & \frac{\partial \phi_1(\vx_{N})}{\partial x_{1}} & \cdots & \frac{\partial \phi_{\npip}(\vx_{1})}{\partial x_{1}}  \\
\vdots & \vdots &  & \vdots \\
\frac{\partial \phi_0(\vx_{N})}{\partial x_{3\na}} & \frac{\partial \phi_1(\vx_{N})}{\partial x_{3\na}} & \cdots & \frac{\partial \phi_{\npip}(\vx_{N})}{\partial x_{3\na}}  
\end{bmatrix} \begin{bmatrix}
 w_0\\
 w_1\\
 w_2\\
  \vdots\\
 w_{\ell}\\
 \vdots\\
 w_{\npip - 1}\\
 w_{\npip}
\end{bmatrix} = \begin{bmatrix}
  y(\vx_{1})\\
  \vdots\\
  y(\vx_{N}) \vspace{0.1cm}\\
  \frac{\partial y(\vx_{1})}{\partial x_{1}} \\
  \vdots \\
  \frac{\partial y(\vx_{1})}{\partial x_{3\na}} \\
  \vdots \\
  \frac{\partial y(\vx_{N})}{\partial x_{1}} \\
  \vdots \\
  \frac{\partial y(\vx_{N})}{\partial x_{3\na}}
\end{bmatrix},\label{eqn:pip_w_grad}
\end{eqnarray}}
where $\na$ is the total number of atoms in the molecule.

From Eq.~\ref{eqn:pip_w_grad}, it becomes evident that there is a need for an efficient and robust framework to compute $\nabla_{\vx} \pipvec$ and forces ($\nabla_{\vx} \mathbf{w}^\top\pipvec$).
Because of this and the wide use of PIP for PESs, there have been previous attempts to make PIP models fully differentiable using both, forward~\cite{pipgrad:bowman:2019} and reverse~\cite{piprevgrad:bowman:2022} mode differentiation.
For $\nabla_{\vx} \pipvec$, a forward mode approach may be memory more efficient, particularly when $\npip$ is large. However, reverse mode differentiation enables the computation of both the energy and force at the same cost as evaluating the energy~\cite{ad:survey}. 
These AD-like frameworks are custom-developed solely for linear PIP models, thus limiting their integration with other regression models and tasks that require higher-order derivatives.

\subsection{\label{sec:anisoPIP}Anisotropic Morse Variables}
One of the foundational components of $\pipvec$ is the Morse variables, $\vgamma$, which are modulated by the length-scale hyper-parameter $\lambda$. Typically, a uniform $\lambda$ is applied across all inter-nuclei distances. Inspired by automatic relevance determination methods in kernel machines \cite{gpbook}, we introduce an anisotropic formulation of the Morse variables, where unique length-scale parameters are assigned to each type of atom-atom distance. 
For example, in ammonia (\ce{NH3}), which contains two different types of distances, \ce{H-H} and \ce{N-H}, the distinct length-scale parameters are $\lambda_{\text{HH}}$ and $\lambda_{\text{NH}}$. 
This approach ensures that each $\lambda_\ell$ parameter exclusively affects its corresponding distance type, maintaining permutational invariance.
Following Eq.~\ref{eqn:piplayer} notation, we define $\pipvecaniso$ as a PIP model with anisotropic Morse variables as,
\begin{eqnarray}
\pipvecaniso(\vx)=(\fpoly \circ \fmono \circ \gamma_{\text{aniso}} \circ d)(\vx),
\end{eqnarray}
where the anisotropic Morse variables are defined by,
\begin{eqnarray}
    \vgamma_{\text{aniso}} &=& \gamma_{\text{aniso}} (\boldsymbol{\lambda}, \boldsymbol{\omega},\vecr) = \exp \left(- \sum_\ell \lambda_\ell  \boldsymbol{\omega}_\ell \odot \vecr \right), \label{eqn:aniso_morse}
\end{eqnarray}
here, $\odot$ denotes the Hadamard product between the distance vector $\vecr$ and the mask vector\footnote{a mask vector is a binary vector used to selectively apply operations to specific elements of another vector or matrix.} $\boldsymbol{\omega}_\ell$, and the sum $\sum_\ell$ is over the unique types of inter-nuclei distances.
The mask vector elements ($\omega_{\ell,j}$) are defined as,
\begin{eqnarray}
 \omega_{\ell, j} = \begin{cases}
 1 & \text{if } r_j \text{ is the same type as } \lambda_\ell, \\
 0 & \text{otherwise}.
\end{cases}    
\end{eqnarray}

In isotropic $\pipvec$ models, the value of $\lambda$ can be optimized with grid search methods or Bayesian optimization\cite{bo:garnett:2023, bo:vargas:2019,bo:vargas:2020}. Instead of relying on these sampling optimization algorithms and following  Ref.~\cite{ad:blondel2022}, we leverage a fully differentiable pipeline and utilize a gradient-based optimization algorithm for the joint optimization of a two-loss function setup,
\begin{eqnarray}
    \vlambda^* &=& \arg\min_{\vlambda \in \mathbb{R}} {\cal L}_{\text{outer}}(\widetilde{\mathbf{w}}(\vlambda)) \nonumber \\
    &\text{subject to}& \;\;\;   \widetilde{\mathbf{w}}(\vlambda) = \arg\min_{\mathbf{w} \in \mathbb{R}} {\cal L}_{\text{inner}}(\mathbf{w},\vlambda), \label{eqn:inner_outer_losses}
\end{eqnarray}
where ${\cal L}_{\text{outer}}$ is the validation loss function, and ${\cal L}_{\text{inner}}$ is the training loss function that jointly depends on $\vlambda$ and $\mathbf{w}$.
Here, $\widetilde{\mathbf{w}}(\vlambda)$ denotes the optimal solution of ${\cal L}_{\text{inner}}$, typically solved using least-squares methods, Eq.~\ref{eqn:pip_w_grad}.
To efficiently solve for $\vlambda^*$, and because of the chain rule, it is essential to determine the gradient $\frac{\partial \widetilde{\mathbf{w}}}{\partial \vlambda}$. 
This gradient can be effectively calculated using implicit differentiation techniques~\cite{ad:vargas:2021,ad:blondel2022}, enabling a systematic and efficient approach to optimizing the anisotropic model parameters for molecular systems with a higher number of unique types of inter-nuclei distances.

% \begin{eqnarray}
%     \vlambda^* = \arg\min_{\vlambda \in \mathbb{R}} \sum_i^N \left ( \widetilde{\mathbf{w}}^\top \pipvecaniso(\vlambda,\vx_{\text{val}}) - y(\vx_{\text{val}}) \right )^2
% \end{eqnarray}

\subsection{\label{sec:pipnn} Permutationally Invariant Polynomial Neural Networks}
Feed-forward neural networks (NNs) known also as multi-layer perceptrons, have been used for PESs\cite{nn:kocer:2022, pipnn:review:carrington:2021}. However, when modeling PESs, neural networks encounter limitations as they do not inherently accommodate the permutation of identical nuclei. Addressing this gap, the integration of PIPs within NNs, coined as PIP-NN ($\fpipnn$), was pioneered by B. Jiang et al. in Ref.~\cite{pipnn:Hua:JCP2013}. This innovative approach significantly enhances neural networks by embedding the necessary symmetry considerations using $\pipvec$ as a symmetrized input to the overall model\cite{pes:shao:2016, invgp:2018}. Following the same framework of function composition, PIP-NNs can be defined as,
\begin{eqnarray}
    \fpipnn(\vx) =  (\fnn \circ \pipvec)(\vx), \label{eqn:pipnn}
\end{eqnarray}
where $\fnn$ can be any NN architecture, commonly feed-forward NNs are the most used NN architecture,
\begin{eqnarray}
    \fnn(\vx) = (g_{L}\circ \cdots \circ g_1 \circ \pipvec )(\vx), \label{eqn:nn}
\end{eqnarray}
where $L$ represents the number of layers, and each layer $g_\ell$ is parametrized by,
\begin{eqnarray}
     g_\ell(\vz) = \sigma(\mathbf{w}_\ell \vz + b_\ell), \label{eqn:act_func}
\end{eqnarray}
where $\sigma$ is the activation function, and $\{\mathbf{w}_\ell,b_\ell\}_{\ell=1}^{L}$ represent the set of parameters for the PIP-NN model.

For PESs, it is a standard practice to train PIP-NNs using both forces and energies jointly,
\begin{eqnarray}
{\cal L}(\boldsymbol{\Theta}, \lambdazero) &=& \frac{1}{N} \sum_i^N \| \nabla \fpipnn (\vx_{i}) - \nabla V(\vx_i) \|_2  \nonumber \\
&&+ \lambdazero \left ( \fpipnn (\vx_{i}) - V(\vx_i)\right)^{2}, \label{eqn:loss_wgrad}
\end{eqnarray}
where the $\lambdazero$ hyper-parameter determines the relative weighting between the energy ($\fpipnn(\vx)$) and force ($\nabla \fpipnn (\vx)$) terms, helping to account for differences in their magnitudes. 
PIP-NNs, like other neural network-based PES models, function as force-field models that train efficiently using stochastic gradient descent methods, with the gradient $\nabla \fnn (\vx)$ computed through automatic differentiation (AD). 

\subsection{\label{sec:pipgp} Permutationally Invariant Polynomial Gaussian Process}
Gaussian Process (GP) models~\cite{gpbook, gp:Csányi:2021,gp:krems:2015:2,gp:vargas:2020} have become one of the most prominent regression models for PESs, and have recently been adapted for larger molecular systems\cite{gppes:krems:2020:2, gppes:Guo:2017,gppes:Kästner:2018,gppes:krems:2016,gppes:krems:2019,gppes:krems:2020,gppes:krems:2022,gppes:krems:2023,gppes:Richardson:2018,gppes:vargas:2018,gppes:vargas:2021,gppes:Zhang:2017}. A GP, denoted as $V(\mathbf{x}) \sim \mathcal{GP}(\boldsymbol{\mu}, \boldsymbol{\Sigma})$, is characterized by its mean ($\mu$) and covariance function ($\Sigma$), which is parameterized by the kernel function $k(\mathbf{x}_i, \mathbf{x}_j)$. The kernel function is a critical component as it describes the similarity between the data points.

A significant advantage of GPs over other probabilistic regression models is their ability to provide a closed-form solution for the posterior distribution with mean and standard deviation functions given by,
\begin{eqnarray}
\mu(\mathbf{x}) &=& k(\mathbf{x}, \boldsymbol{X})^\top \left [ K(\boldsymbol{X}, \boldsymbol{X}) + \sigma_n \mathbb{I} \right ]^{-1} \boldsymbol{y}, \label{eqn:gp_mean}\\
\sigma(\mathbf{x}) &=& - k(\mathbf{x}, \boldsymbol{X})^\top \left [ K(\boldsymbol{X}, \boldsymbol{X}) + \sigma_n \mathbb{I} \right ]^{-1} k(\mathbf{x}, \boldsymbol{X}) \notag \\
&& + k(\mathbf{x}, \mathbf{x}), \label{eqn:gp_cov}
\end{eqnarray}
where for PESs $\boldsymbol{X}$ and $\boldsymbol{y}$ represent the geometric configurations and the corresponding energies. 
In GPs, the kernel function depends on training parameters ($\boldsymbol{\theta}$) that are optimized by maximizing the logarithm of the marginal likelihood (LML),
\begin{eqnarray}
\log p(\boldsymbol{y} | \boldsymbol{X}, \boldsymbol{\theta}) &=& -\frac{1}{2} \boldsymbol{y}^\top \left [ K(\boldsymbol{X}, \boldsymbol{X})  + \sigma_n \mathbb{I} \right ]^{-1} \boldsymbol{y} \label{eqn:gp_lml} \\
&& - \frac{1}{2} \log \left | K(\boldsymbol{X}, \boldsymbol{X}) + \sigma_n \mathbb{I} \right | - \frac{N}{2} \log 2 \pi. \nonumber
\end{eqnarray}
For further details regarding GPs, we refer the reader to Refs.~\cite{gpbook,bo:garnett:2023}.

As discussed in Refs.~\cite{invgp:2018,gpbook}, to construct a GP that accurately represents an invariant function, the kernel must also be invariant. 
Following the approach of C. Qu \emph{et al.} in Ref. \cite{pipgp:vargas:2018}, we define PIP-GPs, as models where $\pipvec$ is used in the kernel function,
\begin{eqnarray}
k(\mathbf{x}_i, \mathbf{x}_j) = k(\pipvec(\mathbf{x}_i), \pipvec(\mathbf{x}_j)), \label{eqn:pip_kernel}
\end{eqnarray}
where $k$ can be any standard kernel such as Radial basis function (RBF) or Matérn, ensuring the covariance matrix $K$ is symmetric and positive semi-definite.
Eq. \ref{eqn:pip_kernel} formulation is kin with the 'Deep Kernel Learning' framework,~\cite{deepkernel} commonly used in contemporary GPs, where NNs parameterize the latent representation of the inputs, followed by the RBF or similar kernel functions.
The training of these GP models can proceed similarly to standard GPs, using gradient-based methods to maximize the LML (Eq. \ref{eqn:gp_lml}). The gradient of the LML with respect to the parameters of $k$ can be computed via automatic differentiation, a common feature in modern GP libraries like GPyTorch\cite{gpytorch:gardner:2018}, GPJax\cite{gpjax:Pinder:2022}, and GAUCHE\cite{gauch}
Additionally, using automatic differentiation, one can compute the derivative of $\mu(\vx)$ with respect to $\vx$ to determine the forces for PESs\cite{gdml:chmiela:2017,pesforce:krems:2021}.

% ----------------------------------------------------------------------------------------
\section{\label{sec:software}Software architecture}
\begin{figure}[h!]
    \centering
    \includegraphics[scale=0.32]{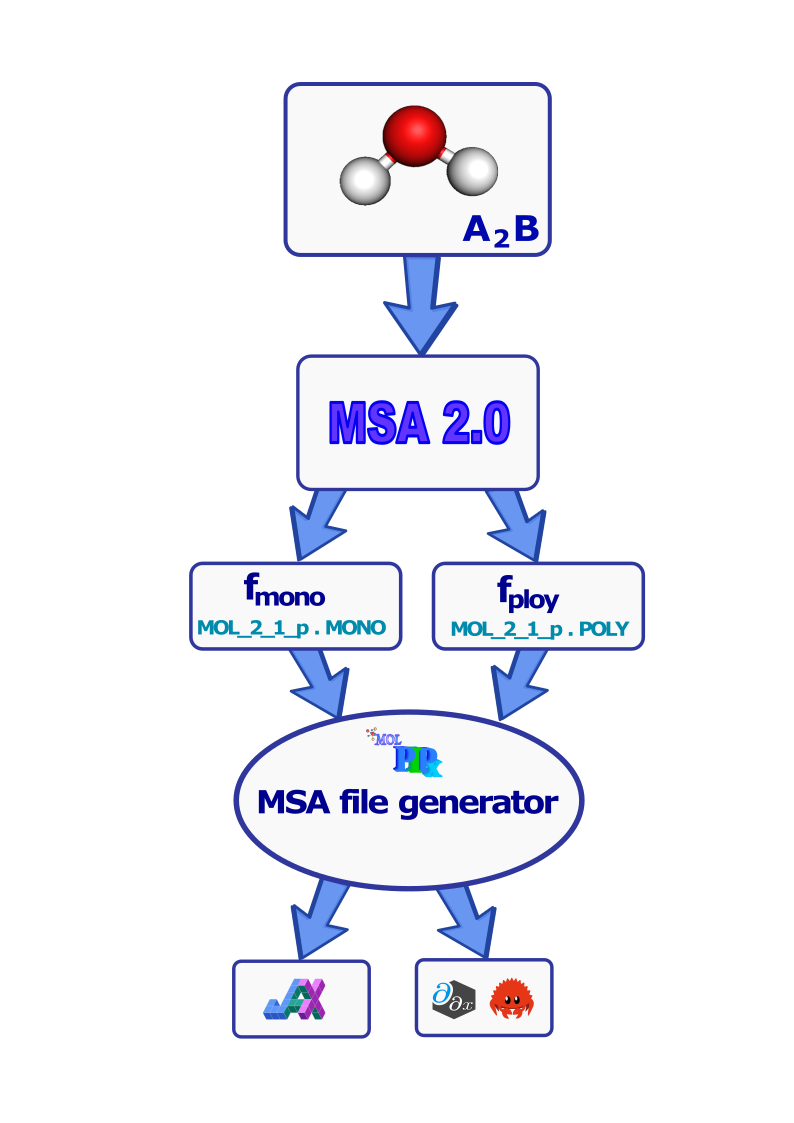}
    \vspace{-20pt}
    \caption{Overview of the workflow to construct the $\fmono$ and $\fpoly$ functions for the selected back-end. The MOLPIPx library translates the \texttt{MOL\_symm\_p.MONO} and \texttt{MOL\_symm\_p.POLY} files generated using the MSA package.
}
    \label{fig:msa_diagram}
\end{figure}

At its core, the library consists of an implementation of the $\fmono$ and $\fpoly$ functions compatible with two automatic differentiation back-ends, e.g., JAX \cite{deepmind2020jax} and EnzymeAD-Rust, and is openly available at \url{https://github.com/ChemAI-Lab/molpipx}. MOLPIPx relies on the \texttt{MSA} package developed by Bowman and co-workers, accessible at \url{https://github.com/szquchen/MSA-2.0}, to generate the monomial and polynomial files, which are then translated into a format compatible with the selected back-end using the \texttt{msa\_file\_generator} function. This workflow is illustrated in Fig.~\ref{fig:msa_diagram} and Listing \ref{lst:msa_file_generator} for the water molecule (\ce{A2B} symmetry). Listings \ref{lst:monomials} and \ref{lst:polynomials} provide examples of the monomial and polynomial functions for the \ce{A2B} molecule in JAX, translated from the MSA using MOLPIPx.
For user convenience, the MOLPIPx library includes pre-configured $\fmono$ and $\fpoly$ functions for the molecules listed in Table \ref{tab:molecules}. The molecules listed in Table \ref{tab:molecules} adhere to a general symmetry notation. For instance, \ce{A3} represents the \ce{H + H2 -> H2 + H} reaction, while water and methane correspond to \ce{A2B} and \ce{A4B}, respectively. These monomial and polynomial functions are also compatible with models based on the many-body expansion.

\begin{table}[ht!]
    \centering
    \caption{List of molecules available in the MOLPIPx Library. For each molecule listed, MOLPIPx has polynomial degrees ranging from 3 to 7.}
    \label{tab:molecules}
    \begin{tabular}{c@{\hskip 0.3in}c}
       \toprule
       \multicolumn{2}{c}{\textbf{Molecule}} \\ \midrule
       A$_{3}$  &  A$_{5}$\\
       A$_{2}$B & A$_{4}$B \\
       ABC &   A$_{3}$B$_{2}$\\
       A$_{4}$ &  A$_{3}$BC\\
       A$_{3}$B & A$_{2}$B$_{2}$C  \\
       A$_{2}$B$_{2}$ & A$_{2}$BCD \\
       A$_{2}$BC & ABCDE \\
       ABCD & \\
       \bottomrule
    \end{tabular}
\end{table}

\begin{lstlisting}[language=Python, label=lst:msa_file_generator,caption= Monomial and polynomial files generator for an \ce{A2B} molecule.]
from molpipx.pip_generator import msa_file_generator
path = './' 
files_msa = 'MOL_2_1_3' # name of the msa files
label = 'A2B_p_3' # name of the output files
# generate mono and poly files in jax
msa_file_generator(files_msa, path, label) 
\end{lstlisting}

% A2B code for the monomials and polynomails
% \onecolumngrid
% \begin{minipage}[t]{0.48\textwidth}
\begin{lstlisting}[language=Python, label=lst:monomials,caption=Monomial function for A$_2$B molecule.]
import jax.numpy as jnp 
from jax import jit
@jit
def f_monomials(r): 
    mono_0 = 1. 
    mono_1 = jnp.take(r,2) 
    mono_2 = jnp.take(r,1) 
    mono_3 = jnp.take(r,0) 
    mono_4 = mono_1 * mono_2 
    return jnp.stack([mono_0, mono_1, mono_2,  
                        mono_3, mono_4]) 
\end{lstlisting}
% \end{minipage}
% \hfill
% \begin{minipage}[t]{0.48\textwidth}
\begin{lstlisting}[language=Python, label=lst:polynomials,caption=Polynomial function for A$_2$B molecule.]
import jax.numpy as jnp 
from jax import jit
from monomial_file import f_monomials as f_mono
@jit
def f_polynomials(r): 
    mono = f_monos(r.ravel()) 
    poly_0 = jnp.take(mono,0) 
    poly_1 = jnp.take(mono,1) + jnp.take(mono,2) 
    poly_2 = jnp.take(mono,3) 
    poly_3 = jnp.take(mono,4) 
    poly_4 = poly_2 * poly_1 
    poly_5 = poly_1 * poly_1 - poly_3 - poly_3 
    poly_6 = poly_2 * poly_2 
    poly_7 = poly_2 * poly_3 
    poly_8 = poly_3 * poly_1 
    poly_9 = poly_2 * poly_5 
    poly_10 = poly_2 * poly_4 
    poly_11 = poly_1 * poly_5 - poly_8 
    poly_12 = poly_2 * poly_6 
    return jnp.stack([poly_0, poly_1, poly_2, poly_3,
                       poly_4, poly_5, poly_6, poly_7, 
                       poly_8, poly_9, poly_10, poly_11,
                         poly_12,]) 
\end{lstlisting}
% \end{minipage}

% \vspace{0.1cm}

% \twocolumngrid
\begin{lstlisting}[language=Python, label=lst:piplayer,caption=Flax module of the PIP layer.]
from typing import Callable
import jax.numpy as jnp
from flax import linen as nn
from molpipx.utils import all_distances, softplus_inverse

@nn.jit
class PIP(nn.Module):
    f_mono: Callable # monomials function
    f_poly: Callable # polynomials function
    l: float = float(1.) # morse variable 
    bias_init: Callable = nn.initializers.constant

    @nn.compact
    def __call__(self, input):
        f_mono, f_poly = self.f_mono, self.f_poly
        _lambda = self.param('lambda',
            self.bias_init(softplus_inverse(self.l)),(1,))
        l = nn.softplus(_lambda)
        d = all_distances(input)  #  distances
        morse = jnp.exp(-l*d)  # morse variables
        return f_poly(morse) # compute PIP vector

@nn.jit
class PIPlayer(nn.Module):
    f_mono: Callable
    f_poly: Callable
    l: float = float(jnp.exp(1))

    @nn.compact
    def __call__(self, inputs):
        # Vectorized version of ``PIP``.
        vmap_pipblock = nn.vmap(PIP, 
            variable_axes={'params': None,},
            split_rngs={'params': False,},
            in_axes=(0,))(self.f_mono, self.f_poly, self.l)
        return vmap_pipblock(inputs)
\end{lstlisting}

One of the main motivations for using JAX as an AD back-end is the existing ML ecosystem. 
For the JAX back-end version, $\pipvec$ is defined as a \texttt{Flax Module dataclass} framework ($\piplayer$), see Listing \ref{lst:piplayer}. Flax also facilitates the incorporation of $\pipvec$ into other ML libraries like GPJax for example. 

To initialize the $\piplayer$, it is required to provide the monomial and polynomial functions in JAX, and the value of $\lambda$, as shown in Listing \ref{lst:piplayer}.
In this library, linear PIP models (Eq.~\ref{eqn:pip_linear}) are defined as $\piplayer$ followed by a \texttt{Dense} layer. The extension for PIP-NN models is straightforward, we alternate an activation function with a \texttt{Dense} layer, as is common practice in ML libraries.
For both linear PIP and PIP-NN models, the optimization of any parameter can be done with gradient-based algorithms. MOLPIPx is compatible with the \texttt{Optax}\cite{deepmind2020jax} and \texttt{JAXopt}\cite{ad:blondel2022} libraries, which contain many gradient descent algorithms commonly used for training modern ML models. 

% RUST back-END
% ----------------------------------------------------------------------------------------
\section{\label{sec:examples}Examples}
The following examples illustrate the use of all previously mentioned regression models, Section \ref{sec:models}, built-in MOLPIPx for the JAX back-end. Currently, the Rust version is limited only to linear models. However, we foresee an easy adaptation to ML libraries based on Rust-AD.  
A more detailed Python code of these examples can be found in \url{https://github.com/ChemAI-Lab/molpipx/tree/main/examples}. 
We use methane, A$_{4}$B, molecule for all examples here.
For all the results presented here, the training data was obtained from Ref.\cite{pipgrad:bowman:2019} and can be downloaded from \url{https://scholarblogs.emory.edu/bowman/potential-energy-surfaces/}.

\subsection{\label{sec:examples_linearpip}Linear Models Training}
As mentioned in Section \ref{sec:linear_pip}, in MOLPIPx the optimization of $\mathbf{w}$ (Eq.~\ref{eqn:pip_linear}) currently uses \texttt{jax.numpy.linalg.lstsq}. 
MOLPIPx provides two additional functions, \texttt{training}, and \texttt{training\_w\_gradients}, the later accounts for the use of $\nabla_{\vx}\pipvec$, Ref.~\cite{pipgrad:bowman:2019}. 
The optimal weights are then copied to a \texttt{Flax}'s parameters \texttt{Pytree} structure, compatible with Flax's \texttt{Dense} layer module (Listing \ref{lst:flaxparam}). 

\begin{lstlisting}[language=Python,label=lst:flaxparam, caption=Linear PIP models.]
from molpipx import PIPlayer, EnergyPIP
from molpipx import training
from molpipx import training_w_gradients, flax_params
# PIP models
pip_model = PIPlayer(f_mono, f_poly) 
e_pip_model = EnergyPIP(f_mono, f_poly)
x0 = jnp.ones([1, na, 3]) # dummy geometry
params_pip = pip_model.init(rng, x0)
params_epip = e_pip_model.init(rng, x0)
# training
X_tr,y_tr = load_data()
w_opt = training(model_pip, X_tr, y_tr)  # Array
params_opt = flax_params(w, params)  # Array to Pytree
\end{lstlisting}
%# prediction
% y = e_pip_model.apply(params_opt,X_new) 

\subsection{\label{sec:examples_value_and_grad}Joint evaluation of energy and force}
For scalar-valued functions, JAX enables the computation of both the function and its gradient simultaneously using the \texttt{jax.value\_and\_grad} function. We have adapted this tool to work with any model developed within MOLPIPx, enabling the joint computation of energy and forces. Listing \ref{lst:energyforce} is a Python code snippet demonstrating the prediction of the energy and forces using the adapted \texttt{get\_energy\_and\_forces} function.
\begin{lstlisting}[language=Python, label=lst:energyforce, caption= Energy and force prediction.]
from molpipx import get_energy_and_forces

# only energy 
y = e_pip_model.apply(params_opt, X) 
# energy and forces
y, f = get_energy_and_forces(e_pip_model.apply,
                                 X, params_opt) 
\end{lstlisting}

\begin{figure}[h!]
    \centering
    \includegraphics[scale=0.4]{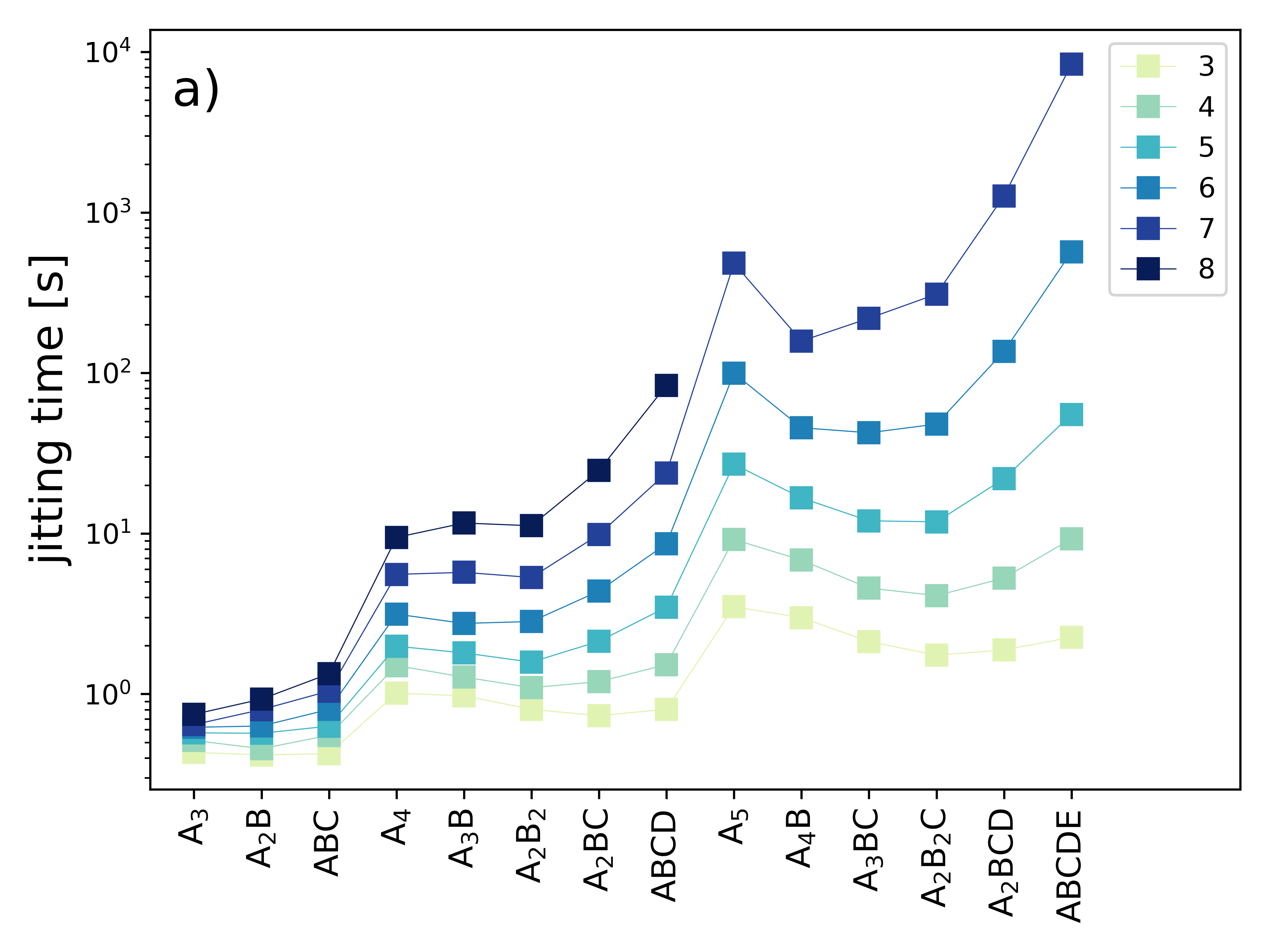}
    \includegraphics[scale=0.4]{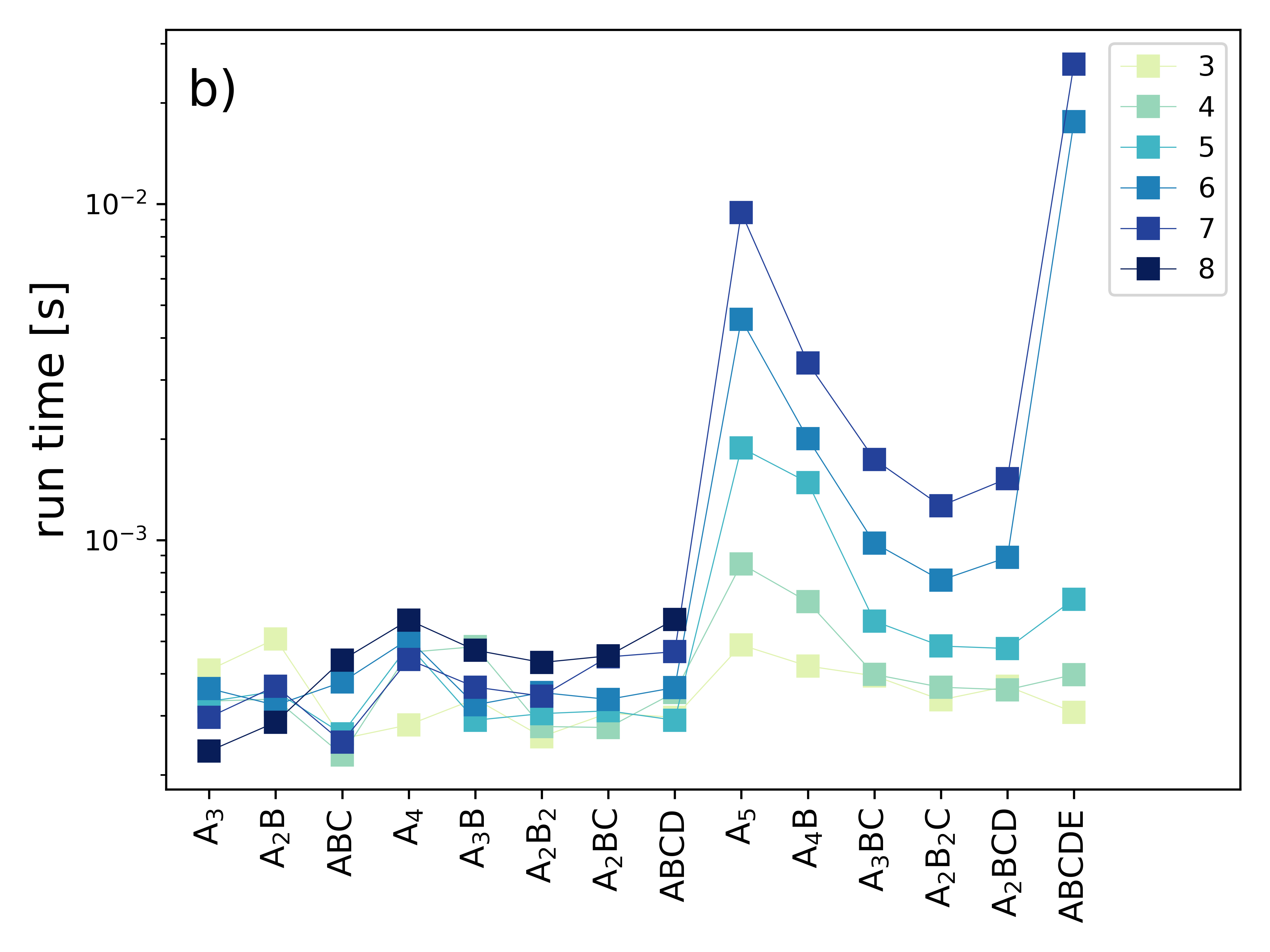}
    \caption{Jitting times (a) and post-jitting run times (b) for the joint computation of energy and force using reverse mode differentiation. Both panels show results for polynomial degrees ranging from 3 to 8. The results for molecules with 5 atoms with $p=8$ are omitted, as the jitting time exceeded 5 hours. 
    All computations were carried in a 2 x Intel(R) Xeon(R) Gold 6248R CPU \@ 3.00GHz with 384GB memory.
    \label{fig:value_and_grad_jit}}
\end{figure}

While all models in MOLPIPx are compatible with \texttt{jax.jit} compilation to enhance performance, this optimization introduces additional computational costs. For the molecules listed in Table \ref{tab:molecules}, we have recorded the time required to jit the \texttt{get\_energy\_and\_forces} function for various linear PIP models with different polynomial degrees. The subsequent prediction times post-jit are shown in Fig. \ref{fig:value_and_grad_jit}. 
As we can observe, the jitting time grows with the complexity of the PIP; however, after jitting, the joint evaluation of the energy and force for $p<4$ is below $10^{-3}s$. The same trend holds using the bigger batches due to the vectorizing map function in JAX.
Both jit and post-jit times for all systems were computed in a 2 x Intel(R) Xeon(R) Gold 6248R CPU 3.00GHz with 384GB memory.
We do not report the jitting times for the JAX version for PIP models with $p=8$ as these are in the order of 10 hours when considering systems with 5 atoms. 
Additionally, we found the compile times for the joint computation of the energy and force using Enzyme-Rust AD in reverse mode for \ce{A5} and \ce{ABCDE}, using a polynomial with degree 8, were 33.5 s and 26 minutes specifically, showcasing a promising speed up compared to the time JAX takes to jit the \texttt{vale\_and\_grad} function. 

\subsection{\label{sec:examples_aniso_opt} Optimization of $\vlambda$ for $\pipvecaniso$}
Here, we will consider the optimization of an anisotropic PIP model for an A$_4$B molecule, where $\vlambda = [\lambda_{\text{AA}}, \lambda_{\text{AB}}]$, and the distance vector is 
\begin{eqnarray}
\begin{split}
\vecr &= 
[\begin{matrix} r_{\text{A}_{1}\text{A}_{2}}, & r_{\text{A}_{1}\text{A}_{3}}, & r_{\text{A}_{1}\text{A}_{4}}, & r_{\text{A}_{1}\text{B}},  & r_{\text{A}_{2}\text{A}_{3}},\end{matrix} \\
 &\qquad\qquad \begin{matrix} r_{\text{A}_{2}\text{A}_{4}}, & r_{\text{A}_{2}\text{B}}, & r_{\text{A}_{3}\text{A}_{4}}, & 
 r_{\text{A}_{3}\text{B}}, & r_{\text{A}_{4}\text{B}} \end{matrix}]
\end{split}
\end{eqnarray}
%$\vecr = [r_{\text{A}_{1}\text{A}_{2}}, r_{\text{A}_{1}\text{B}}, r_{\text{A}_{2}\text{B}}]$. 

%\begin{eqnarray}
%    \vecr = \begin{bmatrix}
% r_{\text{A}_{1}\text{A}_{2}}, & r_{\text{A}_{1}\text{A}_{3}}, & r_{\text{A}_{1}\text{A}_{4}}, & r_{\text{A}_{1}\text{B}},  & r_{\text{A}_{2}\text{A}_{3}}, & r_{\text{A}_{2}\text{A}_{4}}, &\\ r_{\text{A}_{2}\text{B}}, & r_{\text{A}_{3}\text{A}_{4}}, & 
% r_{\text{A}_{3}\text{B}}, & r_{\text{A}_{4}\text{B}} \\
%\end{bmatrix}.
%\end{eqnarray}

For an A$_4$B molecule, the mask vectors ($\boldsymbol{\omega}$) are defined as,
\begin{eqnarray}
    \boldsymbol{\omega} = \begin{bmatrix}
 \boldsymbol{\omega}_{\text{AA}}^\top \vspace{0.1cm} \\ 
 \boldsymbol{\omega}_{\text{AB}}^\top  
\end{bmatrix}  = \begin{bmatrix}
1 & 1 & 1 & 0 & 1 & 1 & 0 & 1 & 0 & 0 \vspace{0.1cm} \\
0 & 0 & 0 & 1 & 0 & 0 & 1 & 0 & 1 & 1 \\
\end{bmatrix}.
\end{eqnarray}

\begin{figure}
    \centering
    \includegraphics[scale=0.43]{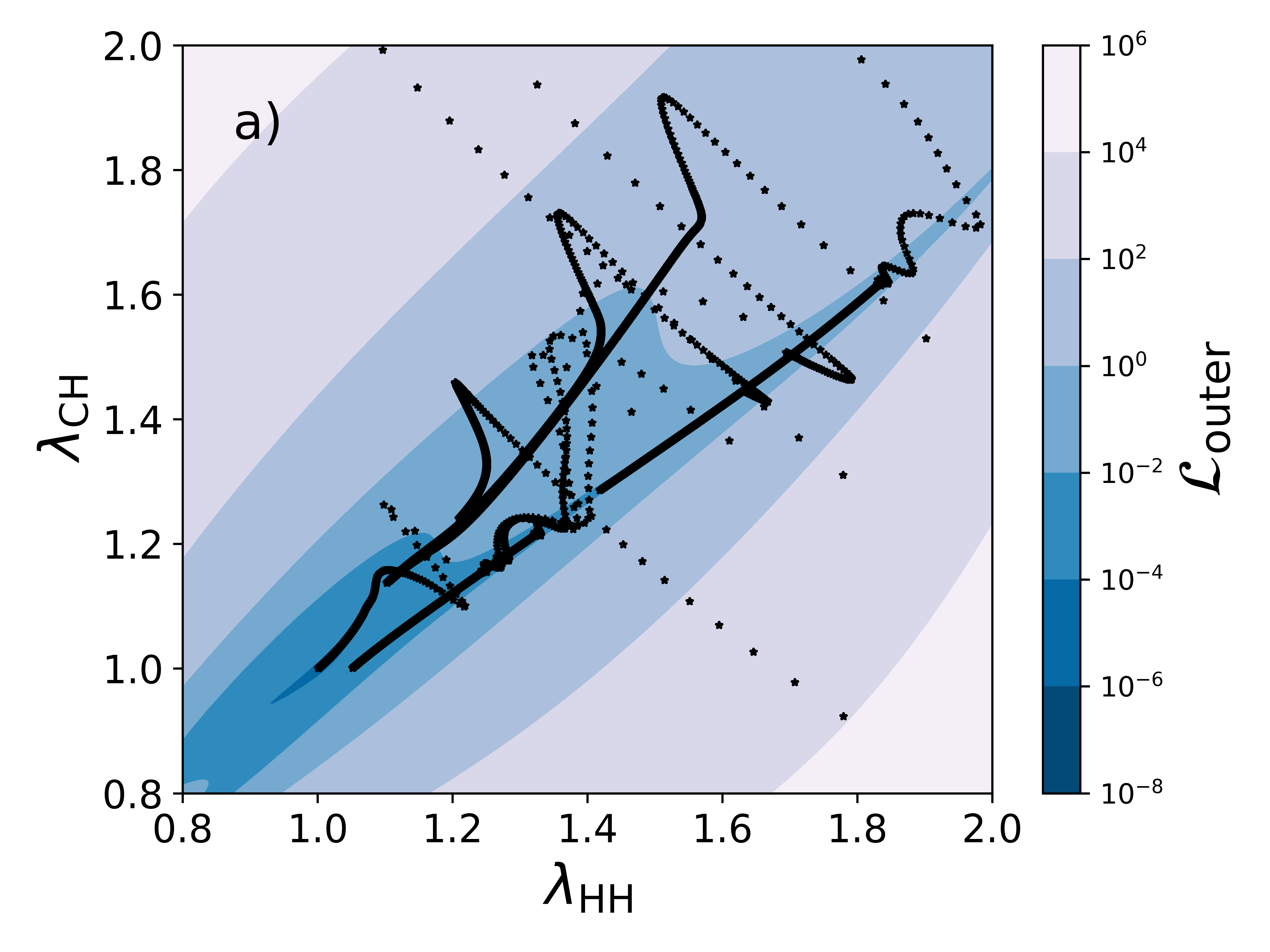} 
    \includegraphics[scale=0.43]{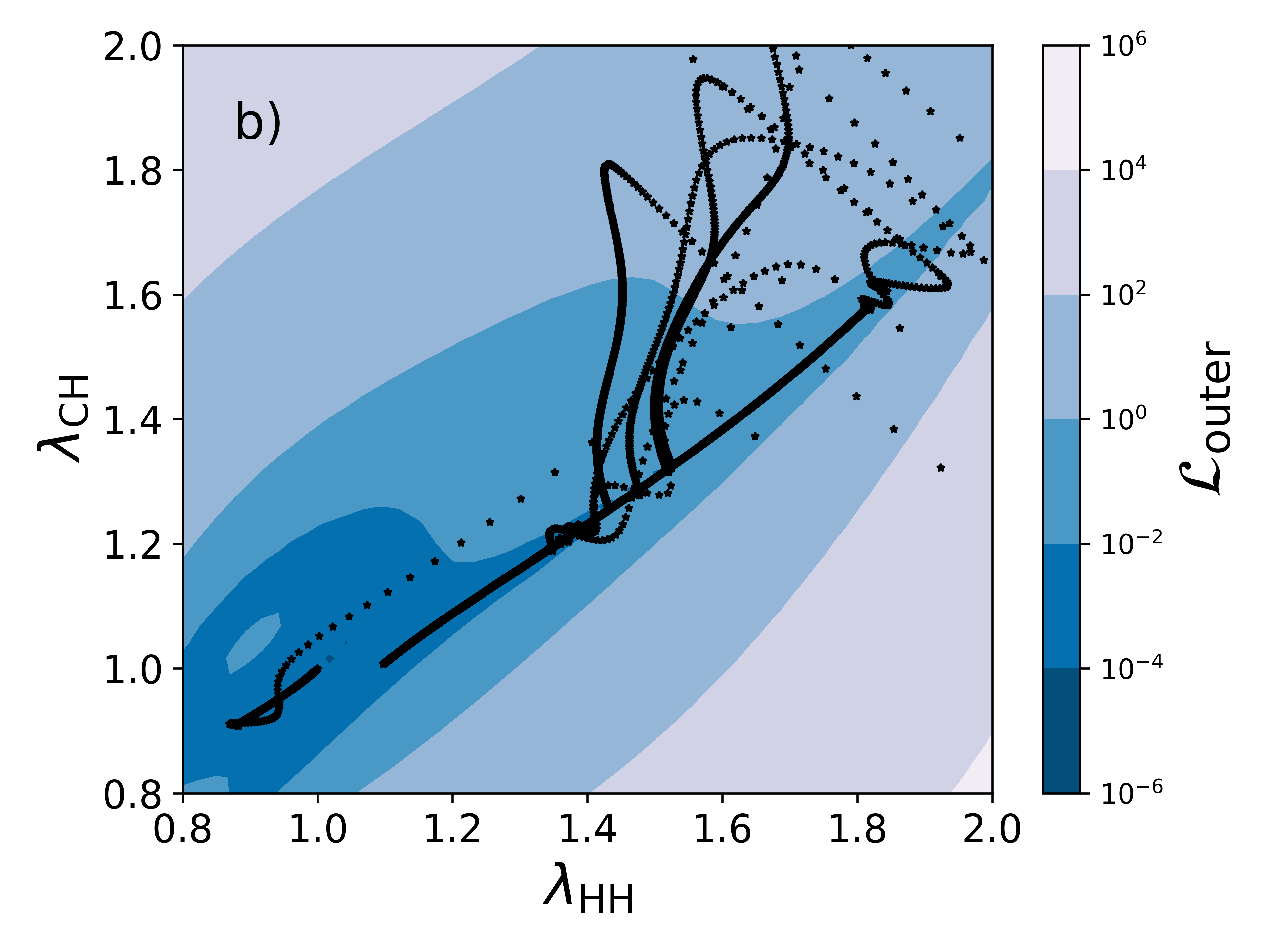} 
    \caption{
    Optimization trajectories of ten random initial $\vlambda$ for methane using an $\pipvecaniso$ model with a third-degree polynomial. Both, the inner and outer loss functions used 500 data points each. (a) shows trajectories optimized with the inner loss function based solely on energy, while (b) incorporates forces. The Adam optimizer was employed with a learning rate of 0.1. \label{fig:anispip_opt}
    }
\end{figure}

As mentioned in Section \ref{sec:anisoPIP}, the optimization of $\vlambda$ depends on the outer (validation) and inner (training) loss functions, defined as,
\begin{eqnarray}
    {\cal L}_{\text{outer}}(\vlambda) &=& \frac{1}{N} \sum_{i}^N \left (\widetilde{\mathbf{w}}(\vlambda)^\top\pipvecaniso(\vx_i,\vlambda) -V(\vx_i) \right )^{2}, \label{eqn:a}
    % {\cal L}_{\text{inner}}(\mathbf{w}) &=& \frac{1}{N} \sum_{i}^N \left ( \mathbf{w}^\top \pipvecaniso(\vx_i,\vlambda) -V(\vx_i) \right )^{2},  
\end{eqnarray}
where the optimal linear parameters, $\widetilde{\mathbf{w}}(\vlambda)$, are found by solving Eq.~\ref{eqn:pip_w_grad}, the minimizer of ${\cal L}_{\text{inner}}$.  
Fig. \ref{fig:anispip_opt} illustrates the optimization trajectories of different randomly initialized $\vlambda$ for \ce{CH4} using a third-degree PIP. We considered two different scenarios, in Fig. \ref{fig:anispip_opt} (a) the inner function only uses energy data, while in Fig. \ref{fig:anispip_opt} (b) we used energy and forces for Eq. \ref{eqn:pip_w_grad}. For both simulations, the outer loss function only depends on energy data.  To ensure the values of $\vlambda$ remain positive, we applied the softplus function.

\subsection{\label{sec:examples_pipnn}Training PIP-NN}
Given the relevance of PIP-NN, we considered two examples, the tunning of the $\lambdazero$ hyper-parameter when jointly considering forces and energies, and the number of neurons' impact on the model's accuracy. 

\begin{figure}[ht!]
    \centering
    \includegraphics[scale=0.45]{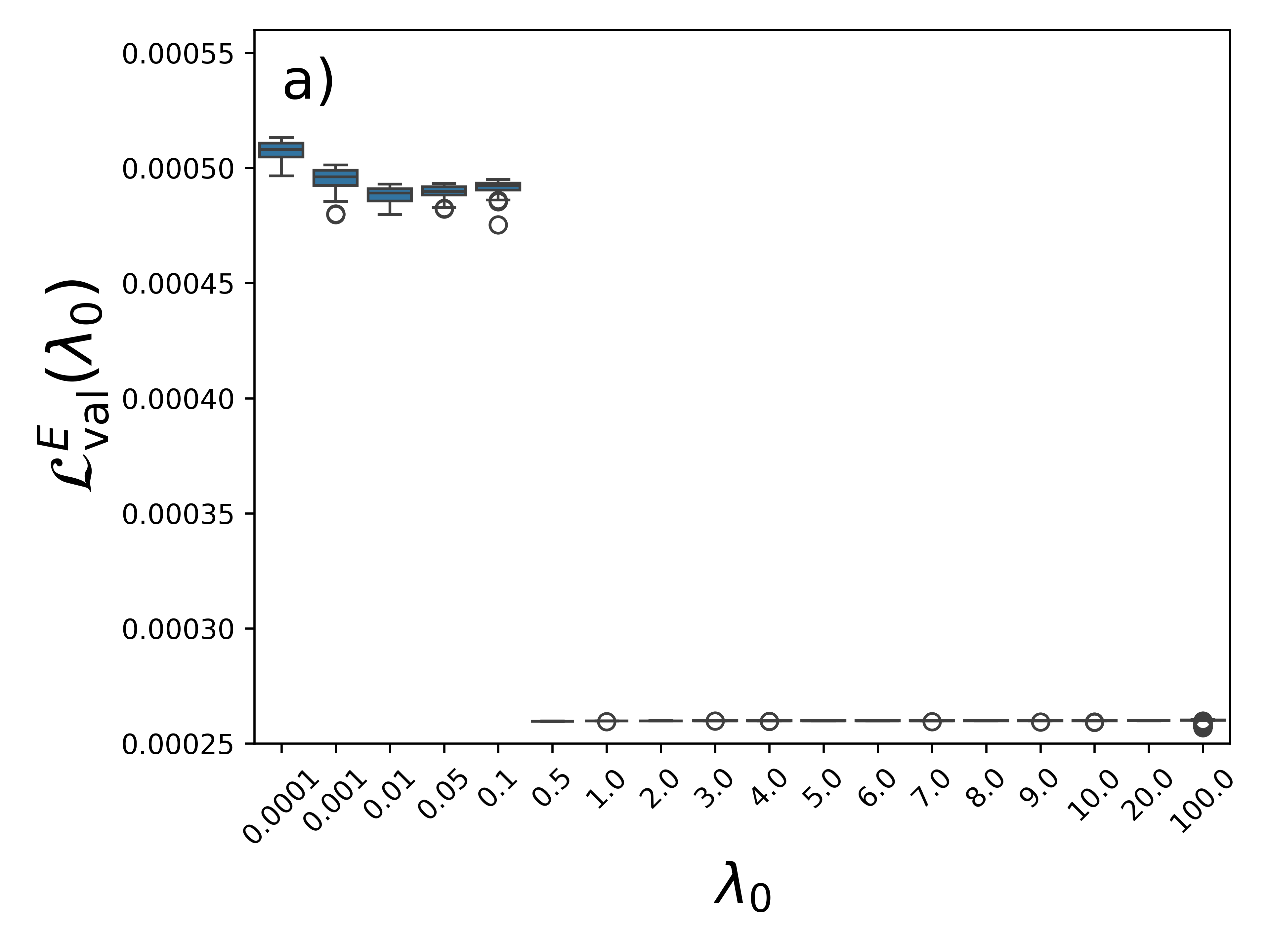}
    \includegraphics[scale=0.45]{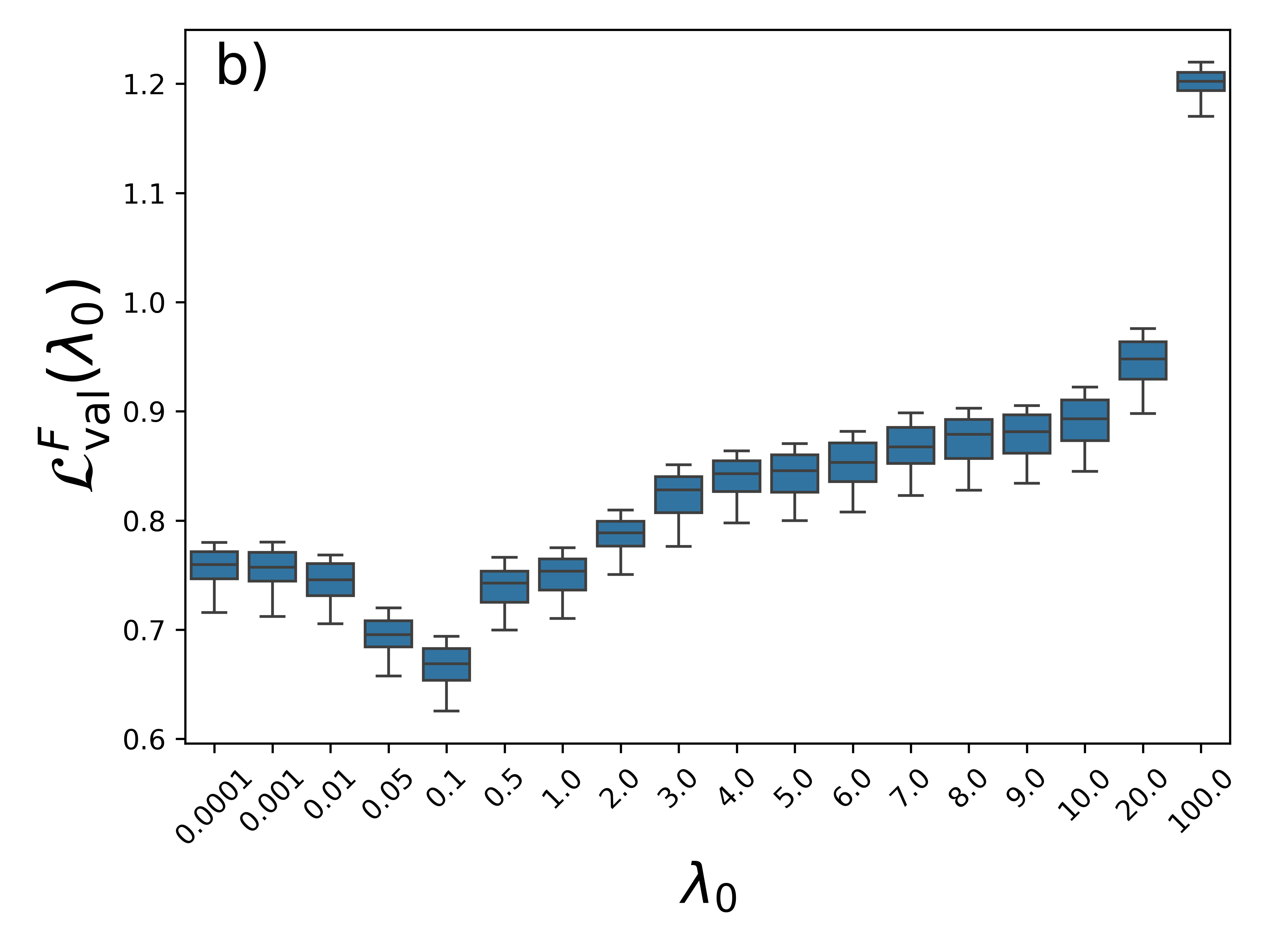}
    \caption{
    (a) Bar plots of the 100 lowest validation errors for energy and (b) forces during training for methane molecule.
    1,000 validation data points are considered to calculate both validation errors. 
    For more details about the PIP-NN architecture and the training see the text. }
    \label{fig:pipnn_lambda0}
\end{figure}

PIP-NN has been widely employed in PESs, as discussed in Section \ref{sec:pipnn}. The training of PIP-NNs can include both energy and forces, as indicated in Eq.~\ref{eqn:loss_wgrad}. Fig. \ref{fig:pipnn_lambda0} depicts the search for the optimal value of the $\lambdazero$ hyper-parameter, using 1,000 training and validation data points. Our grid search protocol demonstrates that the optimal value is $\lambdazero = 0.1$ for a third-degree polynomial PIP-NN with two layers, each containing 128 neurons using the $\texttt{tanh}$ activation function. This value corresponds to the lowest validation error for the forces. The training was conducted using the Adam algorithm with a learning rate of $2\times10^{-3}$ with a batch size of 128.

\begin{figure}[ht!]
    \centering
    \includegraphics[scale=0.45]{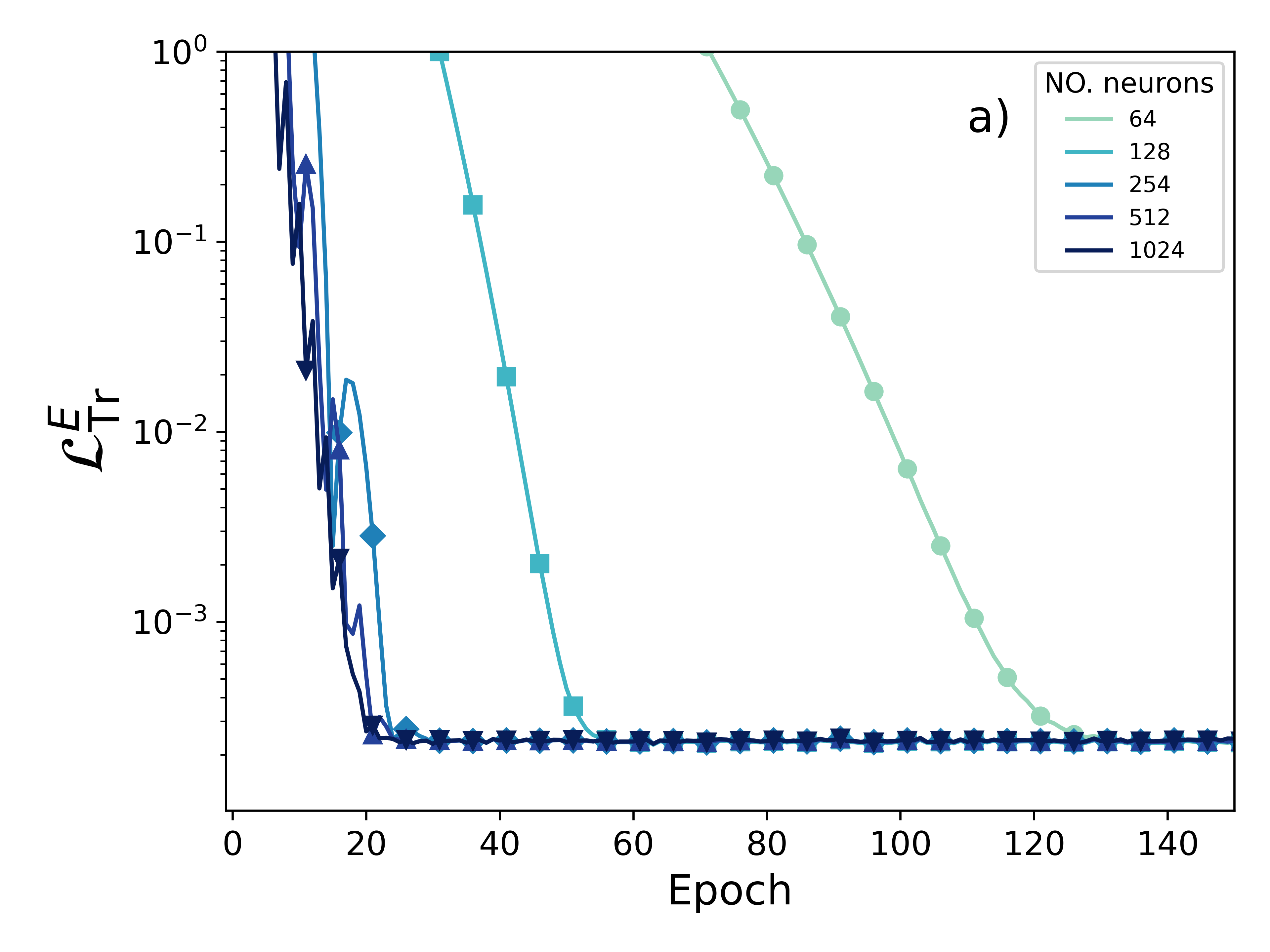}
    \includegraphics[scale=0.45]{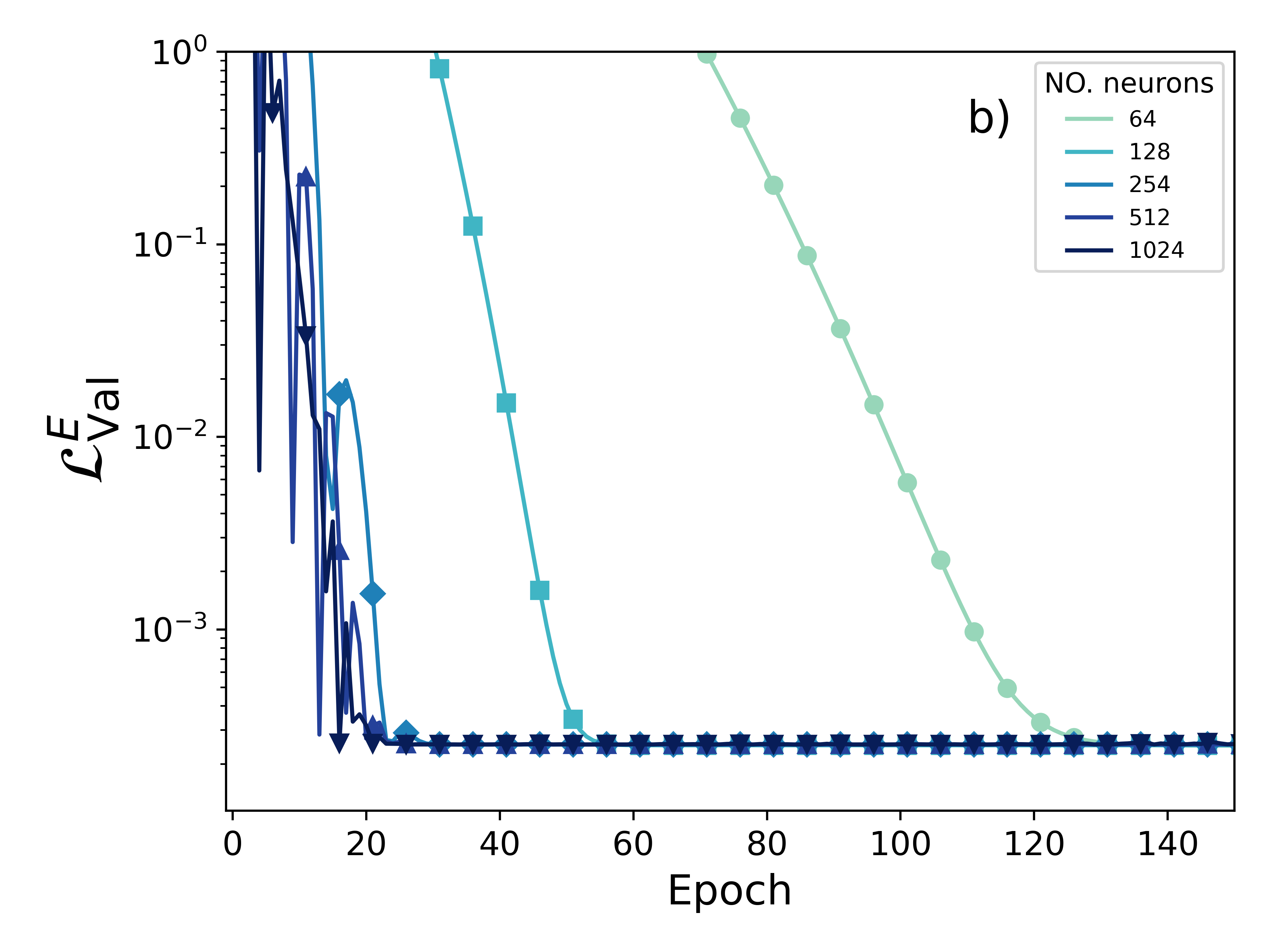}
    \caption{
    (a) Training loss and (b) validation loss  for energy predictions with varying numbers of neurons in a three-layer neural network for the methane molecule. The plots are based on 1,000 training and 5,000 validation data points. }
    \label{fig:pipnn_train_val}
\end{figure}

The training and validation learning curves provide insights into identifying the best-performing model. Fig. \ref{fig:pipnn_train_val} (a) shows the training loss across different epochs for models with varying numbers of neurons in a three-layer architecture, based on 1,000 training energy data points for the methane molecule. Furthermore, the validation loss curve, Fig. \ref{fig:pipnn_train_val} (b), helps assess the model's generalization capability to unseen data. These results indicate that for \ce{CH4}, the best model has 128 neurons, as it smoothly converged more effectively compared to the other models.

%MT results 

\subsection{\label{sec:examples_pipgp}Training PIP-GP}

\begin{figure}
    \centering
    \includegraphics[scale=0.26]{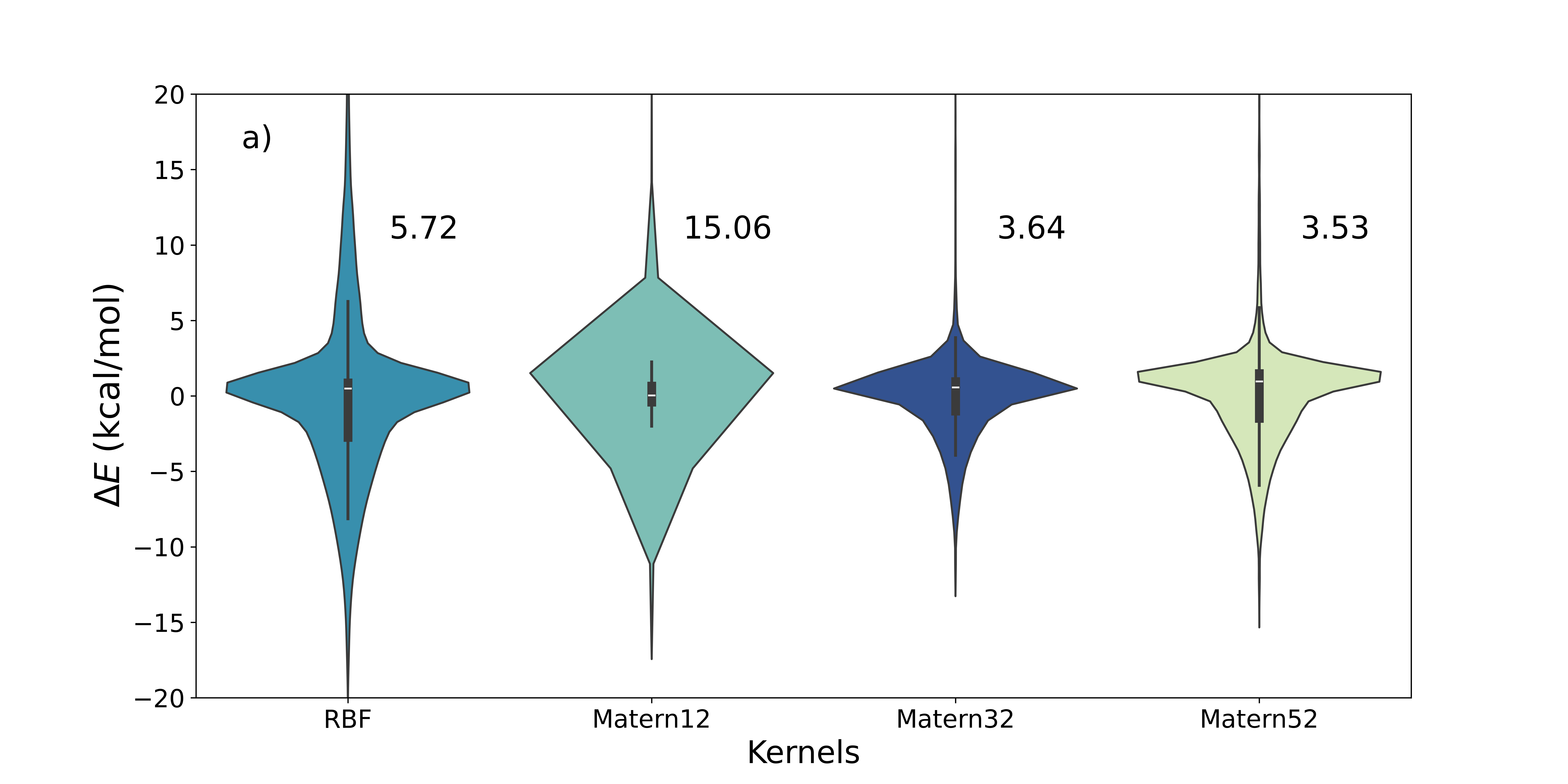} 
    \includegraphics[scale=0.26]{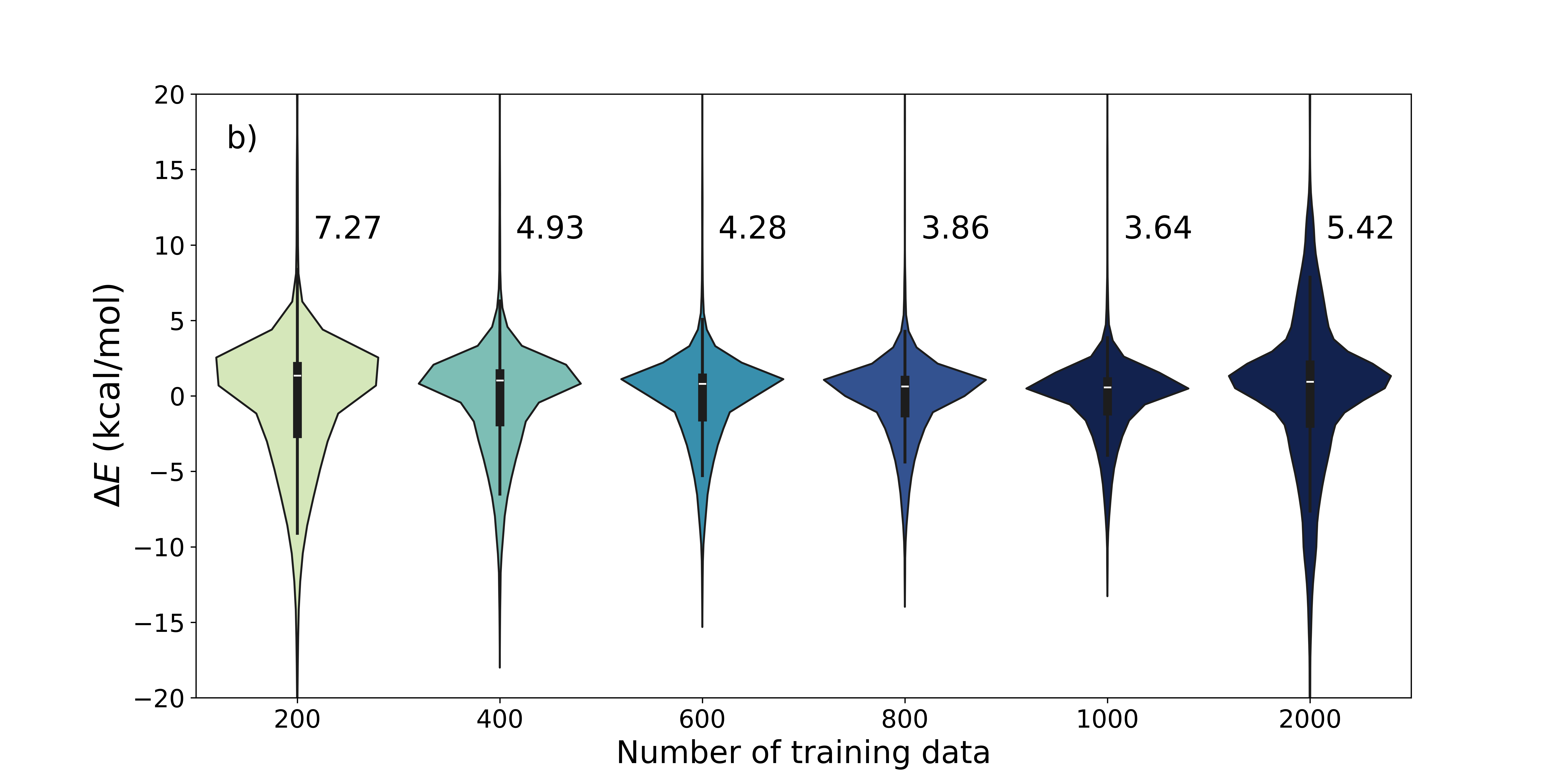} 
    \includegraphics[scale=0.23]{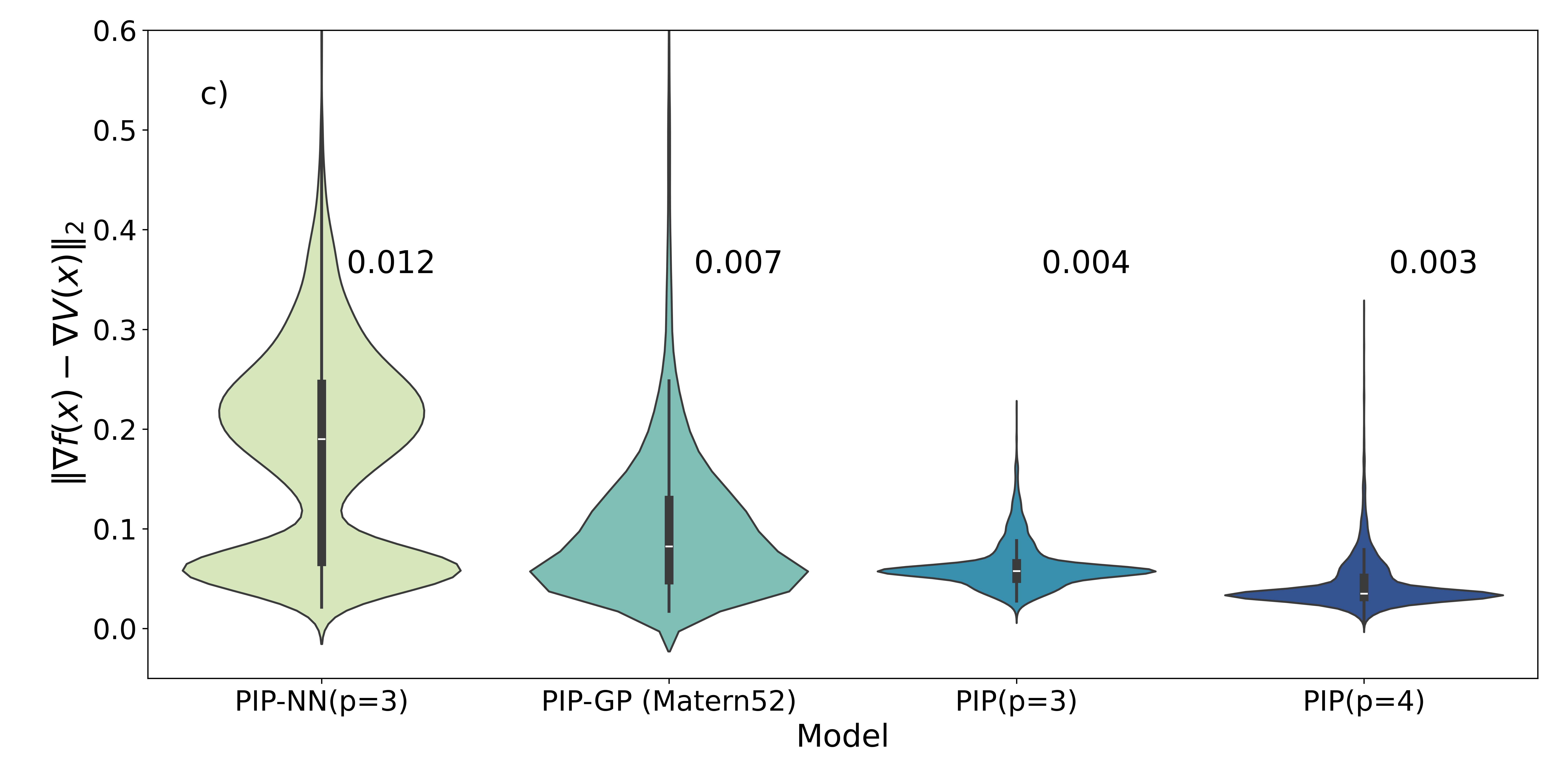} 
    \caption{
    (a) Comparison of different kernel functions for 1,000 training data points. (b) Performance evaluation of the PIP-GP model with the Matern-$\tfrac{5}{2}$ kernel using different training data points. (c) Violin plot of the $\mathit{L}^2$ norm of the difference between the predicted and true forces for different models. The models were trained without force data.  Experiments were carried out on the methane molecule. The numbers represent the RMSE values. All comparisons were performed using 5,000 test data points. \label{fig:kernels}
    }
\end{figure}

As previously discussed, Section \ref{sec:pipgp}, the kernel parameters ($\boldsymbol{\theta}$) in PIP-GPs are optimized by maximizing LML, Eq.~\ref{eqn:gp_lml}. After determining the optimal parameters, the model can predict energy for different coordinates or geometrical configurations. 
% There are several choices of the kernel function for GP. 
Fig. \ref{fig:kernels} shows the results for \ce{CH4} using different kernel functions, with models trained on 1,000 data points and evaluated on 5,000 unseen data points. Based on our results, Fig. \ref{fig:kernels} (a), the Matern-$\tfrac{5}{2}$ kernel demonstrated superior performance with respect to PIP-GP with other kernels. Furthermore, Fig. \ref{fig:kernels} (b) illustrates the impact different number of training data points have on a PIP-GP with the Matern-$\tfrac{5}{2}$ kernel. 

In GPs and PIP-GPs, the mean is tractable, therefore, we can estimate its force analytically too.
Fig. \ref{fig:kernels}(c) shows the $\mathit{L}^2$ norm between the predicted and true forces, for 5,000 unseen data points. The compared models were trained with 1,000 data points only using energy. 
The energy results show that the linear PIP models with $p=3$ and $p=4$ achieve test RMSE values of 1.9 and 1.7 kcal/mol, respectively, which are more accurate than the other PIP models. A similar trend was found for predicting the forces; see Fig. \ref{fig:kernels} (c).

% ----------------------------------------------------------------------------------------
\section{\label{sec:future}Potential extensions}
MOLPIPx is designed as a flexible library for PESs, compatible with existing machine learning frameworks such as Flax and GPJax, as well as future Rust-based libraries. Currently, MOLPIPx relies on pre-existing training data and does not include built-in training or sampling protocols for each PIP-based model. However, it provides users with the flexibility to experiment with various optimization, sampling, and active learning strategies \cite{sampling:jung:2024,sampling:loeffler:2020,sampling:qi:2024}.
In future releases, we plan to include basic sampling and active learning algorithms to improve the overall accuracy of PES models. Additionally, we will integrate MOLPIPx with existing quantum chemistry packages such as PySCF and Psi4, among others. These new features will enable users to enhance both the accuracy and utility of PES models in the study of chemical systems.

Because of the wide use of linear PIP-based models for PESs, we aim to integrate packages like Lineax \cite{lineax2023}, which offer more efficient linear solvers, and to incorporate pruning techniques, such as the L1-Lasso method, which can reduce the number of terms in the PIP vector, accelerating the computation of forces. 
To enhance the PIP-GP framework, we will incorporate gradients directly into the GP model, as demonstrated in previous works \cite{gdml:chmiela:2017,gdml:chmiela:2018,gdml:chmiela:2019}. This approach will allow us to train models using both forces and energies simultaneously, leading to more accurate and robust PES fitting by fully leveraging the available data.

Looking ahead, we plan to extend the package to PyTorch \cite{pytorch:paszke:2019}, given its status as one of the most widely used ecosystems for machine learning models. Additionally, we aim to expand the Rust-based component of the package by developing anisotropic PIPs and a simple feed forward neural networks  tailored for PIP-NN. This library will also ensure compatibility with the \href{https://github.com/sarah-quinones/faer-rs}{faer-rs} Rust-based linear algebra library \cite{faer}, thereby enhancing the versatility and adaptability of the package.

Finally, the monomial and polynomial functions listed in Table \ref{tab:molecules} enable the development of PESs, where PIPs are applied to each $n$-body term in the many-body expansion. This approach was recently employed for hydrocarbons \cite{pip:bowman:2024} (\ce{C14H30}), where the four-body terms in the expansion were \ce{A4}, \ce{A3B}, \ce{A2B2}, \ce{A2BC}, and \ce{ABCD}, as well as in the MB-pol many-body potential for water \cite{pip:babin:2014, pip:moberg:2016}. MOLPIPx could further benefit from using mask vectors, such as those utilized in the anisotropic Morse variables described in Section \ref{sec:anisoPIP}, and from the development of training schemes based on gradient-based or least-squares methods.

% ----------------------------------------------------------------------------------------
\section{\label{sec:summary}Summary}
MOLPIPx offers a comprehensive and adaptable platform that integrates PIPs with cutting-edge machine learning libraries by leveraging modern computational tools such as JAX and Enzyme-AD. Due to its end-to-end differentiable nature, MOLPIPx supports a range of regression models for PESs, from traditional linear methods to advanced models like neural networks, Gaussian Processes, and anisotropic Morse variables. Furthermore, within MOLPIPx, the computation of forces is performed automatically for any proposed PIP-based model.

We provide the monomial and polynomial functions for fifteen different molecular systems, however, MOLPIPx can be interfaced with the MSA package to translate those files into pure Python functions using JAX as the main numerical back-end library. 
Finally, we also provide a comprehensive set of examples for each regression model to illustrate the use of MOLPIPx.

\section{Acknowledgement}
The authors thank Chen Qu for fruitful discussions. MSD is thankful for the Rust Foundation's support. 
This research was partly enabled by support from the Digital Research Alliance of Canada and NSERC Discovery Grant No. RGPIN-2024-06594.

% \appendix

% \section{Appendixes}

\nocite{*}
\section*{References}
\bibliography{references}% Produces the bibliography via BibTeX.

\end{document}